\documentclass[aps,prd,reprint,groupedaddress,nofootinbib]{revtex4-1}

\usepackage{graphicx}
\usepackage{longtable}

\usepackage{relsize}

 \usepackage{amssymb} \usepackage{amsmath}
\usepackage{color}
\usepackage{url}
\usepackage[bookmarks, breaklinks, colorlinks,urlcolor=black, citecolor=red, 
linkcolor=blue]{hyperref}

\usepackage{tikz}
\usetikzlibrary{positioning,arrows}
\usetikzlibrary{decorations.pathmorphing}
\usetikzlibrary{decorations.markings}

\usepackage{subfigure}

\def\beq{\begin{equation}}
\def\eeq{\end{equation}}
\def\barr{\begin{array}}
\def\earr{\end{array}}
\def\dis{\displaystyle}

\usepackage{cancel}
 \usepackage[normalem]{ulem}
\usepackage{color}

\definecolor{darkcyan}{cmyk}{1,0,0,0.4}

\definecolor{darkred}{cmyk}{0,1,1,0.4}

\def\lapp{\mathrel{\rlap{\raise.5ex\hbox{$<$}}
                    {\lower.5ex\hbox{$\sim$}}}}
\def\gapp{\mathrel{\rlap{\raise.5ex\hbox{$>$}}
    {\lower.5ex\hbox{$\sim$}}}}
\def\kev{\, {\rm keV}}
\def\mev{\, {\rm MeV}}
\def\gev{\, {\rm GeV}}
\def\xenon{{\sc xenon1t} }
\def\Babar{{\mbox{\slshape B\kern-0.1em{\smaller A}\kern-0.1em B\kern-0.1em{\smaller A\kern-0.2em R}}}}
\begin{document}

\tikzset{
  scalar/.style={dashed,draw=black, postaction={decorate},decoration={markings,mark=at position .5 with {\arrow[draw=black,thick]{>}}}}
}

\title{Dark Matter, Muon Anomalous Magnetic Moment and the XENON1T Excess}

\author{Debajyoti Choudhury}
\email{debajyoti.choudhury@gmail.com}
\affiliation{Department of Physics and Astrophysics, University of Delhi, Delhi 110007, India}
\author{Suvam Maharana}
\email{msuvam221@gmail.com}
\affiliation{Department of Physics and Astrophysics, University of Delhi, Delhi 110007, India}
\author{Divya Sachdeva}
\email{divyasachdeva951@gmail.com}
\affiliation{Department of Physics, Indian Institute of Science Education and Research Pune, Pune 411008, India}
\author{Vandana Sahdev}
\email{vandanasahdev20@gmail.com}
\affiliation{Department of Physics and Astrophysics, University of Delhi, Delhi 110007, India}

%\date{\today}

\begin{abstract}
  A very economic scenario with just three extra
  scalar fields beyond the Standard Model is invoked to explain the
  muon anomalous magnetic moment, the requisite relic abundance of
  dark matter as well as the \xenon excess through the inelastic
  down-scattering of the dark scalar.

  \end{abstract}
\maketitle

%%%%%%%%%%%%%
%%%%%%%%%%%%%******************************************
%%%%%%%%%%%%%
\section{Introduction}
\vspace*{-0.2cm} The observation of an excess in the electronic recoil
events at the \xenon detector~\cite{Aprile:2020tmw} has elicited much
activity, especially in the context of Dark Matter (DM). A multitude
of explanations have been proposed, incorporating different
  mechanisms, such as boosted DM
\cite{Kannike:2020agf,Cao:2020bwd,*Primulando:2020rdk,*Alhazmi:2020fju,*DelleRose:2020pbh,*Ko:2020gdg,*Fornal:2020npv},
inelastic DM-target scattering
\cite{Harigaya:2020ckz,*Baryakhtar:2020rwy,*Bramante:2020zos,*Baek:2020owl,*Chao:2020yro,*An:2020tcg,*Shakeri:2020wvk,*Lee:2020wmh}
as well as many others
\cite{Smirnov:2020zwf,*Takahashi:2020bpq,*Alonso-Alvarez:2020cdv,*Su:2020zny,*Du:2020ybt,*Chen:2020gcl,*Bell:2020bes,*Paz:2020pbc,*Nakayama:2020ikz,*Gelmini:2020xir,*Jho:2020sku,*Zu:2020idx,*Okada:2020evk,*Dey:2020sai,*Choi:2020udy}.
  The very structure of the excess demands that not only the DM
  particle be relatively light, but also that the recoil energy
  satisfy $1\kev \lapp E_{\rm rec.} \lapp 5\kev$. To reconcile such a
  DM with the correct relic abundance and yet survive cosmological
  constraints emanating from large-scale structure formation, big-bang
  nucleosynthesis, cosmic microwave background~\cite{Berger:2016vxi},
  supernovae~\cite{Chang:2018rso} etc., has been a herculean task.  In
  this paper, we point out that a relatively simple model can not
  only satisfy all such constraints but also successfully address
  another long-standing issue that the Standard Model (SM) faces,
  namely an explanation of $a_\mu$, the anomalous magnetic moment of
  the muon. Furthermore, it promises exciting signals at currently
  operating experiments.
\vspace*{-0.4cm}
\section{Model}
\vspace*{-0.2cm}
Eschewing the more common fermionic DM, we consider the simpler
alternative, {\em viz.}  a complex scalar field $\phi$. The lack of
excess events in the first bin at
\xenon~\cite{Aprile:2020tmw,Kannike:2020agf} restricts $m_\phi \lapp
\mathcal{O}(1)\gev$.  The dark sector communicates with the SM
particles through a light real scalar field $\omega$ which also serves
to generate a contribution to $a_\mu$.  There are some advantages to
choosing a scalar mediator as opposed to the more popular dark
photon. For one, it is the most economic construction in terms of
field content. Secondly, $\omega$ can both be the mediator as well as
potentially engender the mass split required for down-scattering. And,
finally, having a dark photon generate a substantial $a_\mu$ would
need it to couple to neutrinos as well (at least in the simpler
constructions) thereby rendering the heavier component of the DM
unstable on cosmological time scales.

Given the field content, the most general scalar potential
has many parameters. For the sake of brevity, consider
here the relevant part of the same, {\em viz.}
%%%%%%
\beq
\barr{rcl}
 V_{\phi, \omega}  & \ni &\dis \mu_\phi^2 \, \phi^*\phi
    + (\Delta^2 \phi^2 + H.c.)\\[1ex]
    &+ & \dis
    2 \mu_{\phi \omega} \phi^* \phi \,  \omega +
          \left[ (A_1 + i A_2) \, \phi^2 + H.c) \right]
         \omega \ .
\earr
    \label{eq:scal_lagr}         
\eeq
%%%%%%
Also possible are other cubic and quartic terms, including, possibly a
$\omega^3$ one. However, unless their coefficients are large, such
terms would not be germane to the issues at hand. While ensuring that
$\phi$ represent a viable DM demands that its classical value (vev) 
vanishes identically, we impose an identical (simplifying) condition
for $\omega$ as well. These conditions and the lightness of the
scalars can be easily achieved by suitably adjusting the parameters of
the full potential.

The presence of the $\Delta^2$ term serves to split the two
components of $\phi \equiv (\phi_2 + i \phi_1) / \sqrt{2}$. For real $\Delta^2$
(an imaginary component to $\Delta^2$ does not change anything qualitatively
beyond introducing an immaterial mixing), the masses are given by (without loss of generality, $\Delta^2 > 0$)
%%%
\beq
m_{2,1} = \sqrt{\mu_\phi^2 \pm 2\Delta^2}  =
\mu_\phi \pm \delta_m  \ , \qquad \delta_m \simeq \Delta^2/\mu_\phi \ .
\label{eq:the_masses}
\eeq
%%%
As we shall see later, for a successful explanation of the \xenon
signal, we would require $\mu_\phi$ to be a few GeVs at best (with a
few hundred MeVs being preferred). In particular, a much heavier DM
would result in too large (and wide) an excess in the low
recoil-energy bins. Furthermore, $\delta_m \sim 2 \kev$ is motivated
not only by the profile of the excess but also by the functional
dependence of the atomic excitation factor, as would be seen in
Section III.
 
The soft trilinear terms in eq.(\ref{eq:scal_lagr}) engender couplings
of the form $g_{ij} \phi_i \phi_j \omega$, with
%%%%%
\beq
g_{11} = \mu_{\phi\omega} - A_1 \ , \quad
g_{22} = \mu_{\phi\omega} + A_1 \ , \quad
g_{12} = -2 A_2 \ .
\label{eq:triple_scalar}
\eeq
%%%%%
While the $g_{ij}$s play nearly equivalent roles in determining the
relic density, $g_{12}$ is key to explaining the \xenon excess.  Also
note that $\langle\omega\rangle \neq 0$ would generate $\Delta^2$.

The messenger $\omega$ can have renormalizable interaction terms with only
the Higgs field with the $H^\dagger H \omega^2$ term
constrained by the limits on the
invisible decay width of $H$. Similarly, a $\omega H^\dagger H$
term would induce a mixing, and is constrained by $B$ and $K$ decays~
\cite{Bird:2004ts,OConnell:2006rsp,Pospelov:2007mp,Batell:2009jf,Aaij:2015tna,Krnjaic:2015mbs}. However, we do not delve into this and focus, instead,
on a leptophilic $\omega$, coupling to fermions through 
dimension-five operators such as~\cite{Liu:2018xkx}
%%%%
\[
    {\cal L}_{\rm int} \ni
    (\omega/\Lambda) H \, \left[ \tilde y_\mu \bar L_2 \mu_R
      + \tilde y_e \bar L_1 e_R \right] + H.c. \ ,
\]
%%%%%
where $L_{1,2}$ are the electron and muon doublets and $\Lambda$ is
the cutoff scale, presumably in the multi-TeV range.  (While $\tilde y_\tau$
could exist as well, it 
does not largely concern us, and we shall remark on its
consequences later.) On symmetry breaking, these
lead to effective Yukawa terms
%%%%
\beq
    {\cal L}_{\rm Yuk}^{\rm eff} \ni 
    \omega \left[ y_{\mu} \bar \mu \mu
      + y_{e}\bar e e \right]  \ , \qquad y_\ell \equiv \tilde y_\ell v /\sqrt{2} \Lambda \ .
\label{eq:eff_Yuk}
\eeq
%%%%%
We shall, henceforth, parametrize
\beq
y_e = n_s \, (m_e / m_\mu) \, y_\mu
   \label{eq:yuk_scaling}
\eeq
where the scaling factor $n_s = {\cal O}(1)$. 

Before delving into phenomenological consequences, we must discuss the
decays. While $\phi_1$ is absolutely stable, $\phi_2$ decays may
  occur at one-loop. Owing to the tiny $\delta_m$, these are
restricted to $\phi_1 + N_1 \gamma + N_2 (\nu\bar\nu)$ alone. 
While $\phi_2 \to \phi_1 + \gamma$ is ruled out from
  considerations of angular momenta, decays into neutrinos are highly
  suppressed owing to the $W$-mass. The leading decay mode,
viz. $\phi_2 \to \phi_1 \gamma \gamma$ proceeds through
the effective $\omega\gamma\gamma$ vertex, with the decay width being
given by
\begin{eqnarray}
\Gamma_{\phi_1\gamma\gamma}\,&=&\,\left[\frac{\alpha\,g_{12}}{3\pi m_{\omega}^2}\sum_{f=e,\mu} \frac{Q_f^2 y_f}{m_f}\right]^2\frac{\delta_m^7}{3360\pi^3m_{2}^2}\nonumber\\
&=&\,4.3\times10^{-31}{\rm s}^{-1}\left[\frac{g_{12}}{10^{-2} {\rm GeV}}\right]^2\left[\frac{0.07  {\rm GeV}}{m_{\omega}}\right]^4 \nonumber \\
&\times&(n_s+1)^2\left[\frac{y_\mu}{10^{-4}}\right]^2\left[\frac{\delta_m}{2 {\rm keV}}\right]^7\left[\frac{0.5 {\rm GeV}}{m_{\phi}}\right]^2
\end{eqnarray}
It might seem that the consequent emission should have been seen by
X-ray observatories such as Chandra or XMM\mbox{-}Newton
etc.\cite{Boyarsky:2007,Boyarsky:2008}. However, note that, with the decay being a 3-body
one, the excess would be a continuum ranging upto $\sim 1$~keV with a
maximum at $\sim 0.8$~keV. The absence of a sharp line naturally
reduces the sensitivity as compared to, say, that in
Ref.~\cite{Bazzocchi:2008}, and even with a conservative
interpretation, the parameter space to the left of the peak in
Fig.~\ref{gphi1} is consistent with the absence of an excess.

On the other hand, when allowed, the partial
widths of $\omega$ into leptonic and scalar channels are given,
respectively, by (here, $\beta_X \equiv (1 - 4 m_X^2 / m_\omega^2)^{1/2}$)
\beq
\Gamma_{\ell\ell}  = y_\ell^2\, m_{\omega} \beta_\ell^3 /(8 \pi) ,
\,
\Gamma_{ij} = g_{ij}^2 \beta_\phi \left( 2 - \delta_{ij} \right)/(32 \pi m_{\omega}).\label{eq:partial_widths}
\eeq
%%%%%%%%%%%%%
%%%%%%%%%%%%%******************************************
%%%%%%%%%%%%%
\vspace*{-1cm}
\noindent
\subsection{Anomalous magnetic moment of the muon}
\begin{figure}
\centering
\includegraphics[width =9cm,height=4.6cm]{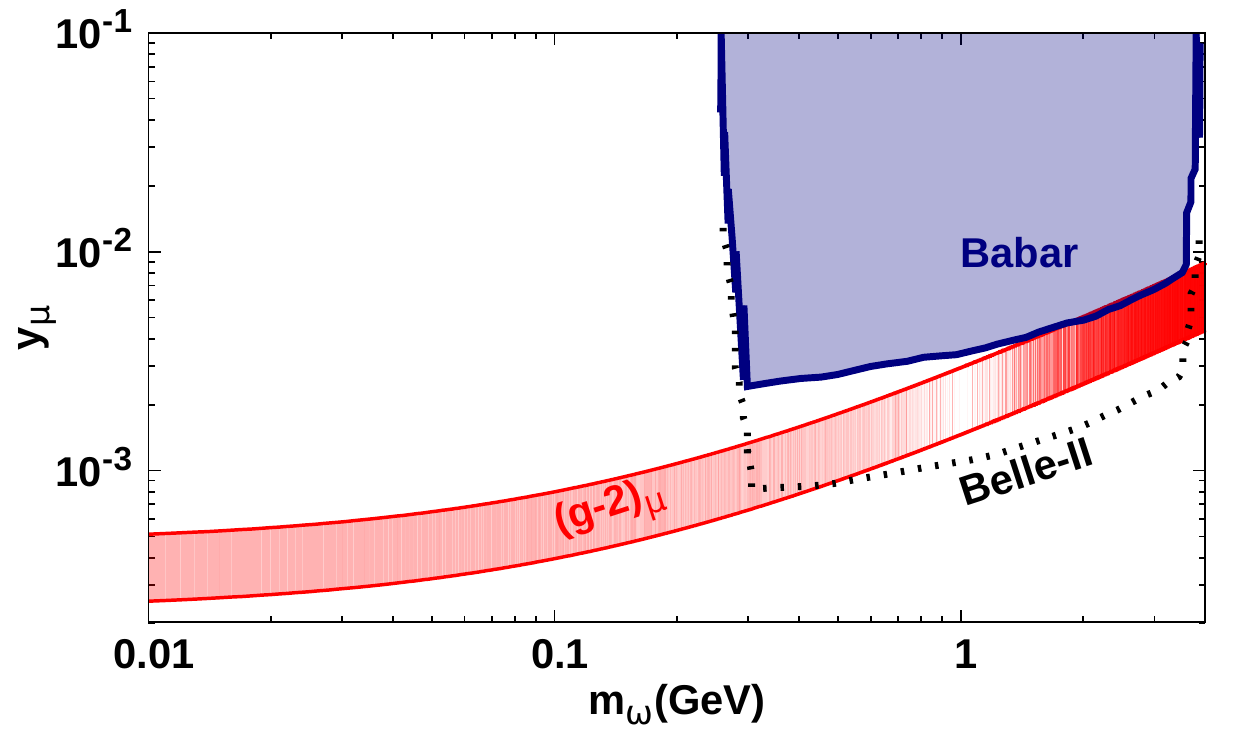}
\vskip -10pt
\caption{\em The $2\sigma$ band favoured by $(g-2)_\mu$ and 
 the constraint from the $4\mu$ final state assuming
  $Br(\omega \to \mu^+\mu^-)= 1.0$. The dotted curve is the projection from 
  BELLE-II experiment~\cite{Batell:2017kty}.}
\vskip -10pt
\label{muoncon}
\vskip -5pt
\end{figure}

The interaction
of eq.(\ref{eq:eff_Yuk}) generates an additional  contribution $\delta a_\mu$
at one-loop itself. The expression is
straightforward~\cite{Leveille:1977rc,Chakraverty:2001yg,Giudice:2012ms,Batell:2016ove,Liu:2018xkx} {\em viz.},
\beq
\delta a_\mu =  \dis \frac{y_\mu^2 m_\mu^2}{8 \pi^2} \,
             \int_0^1 \frac{z^2 \, (2 - z) \, dz}{m_\omega^2 (1 - z) + m_\mu^2 z^2}
\label{eq:a_mu}
\eeq
and explaining the discrepancy~\cite{Tanabashi:2018oca}
\[
\delta a_\mu \equiv a_\mu^{\rm exp} - a_\mu^{\rm SM}
= (261 \pm 63 \pm 48) \times 10^{-11} \ ,
\]
yields a band in the $y_\mu$-$m_{\omega}$ plane (Fig.\ref{muoncon}).

Even for $y_e = 0$, the \Babar\ search for dark photons in the $4\mu$ final
state~\cite{TheBABAR:2016rlg} can be reinterpreted in terms of
$e^+ e^- \to \mu^+ \mu^- \omega$ followed by $\omega \to \mu^+ \mu^-$ 
~\cite{Batell:2016ove,Batell:2017kty,Liu:2018xkx} for the scalar
mediator yielding an upper bound of $y_{\mu} \lapp (2-8)\times 10^{-3}$
for 0.2 GeV $\lapp m_{\omega} \lapp 3$ GeV. As can be easily ascertained
from Fig.\ref{muoncon}, this has little bearing on the solution for $(g-2)_\mu$.

Owing to its much smaller size, a non-zero
$y_e$ does not materially affect this conclusion. The situation,
though, could change drastically if a $y_\tau $ were to exist, for it
would lead to $e^+ e^- \to \tau^+ \tau^- \omega \to \tau^+ \tau^-
\ell^+ \ell^-$ at \Babar~\cite{BABAR:2020oaf}.  In the event of $y_i
\propto m_i$, the constraints on $y_\tau$ could be interpreted in
terms of much stronger bounds on $y_\mu$ and $y_e$. Note,
however, that both the \Babar\ analyses assume $Br(\omega \to \mu^+ \mu^-)=
1.0$ for $m_\omega\,>\,0.21\gev$ or $Br(\omega \to e^+e^-)= 1.0$ for
$0.04\gev\,<\,m_\omega\,<\,0.21\gev$. In our scenario, whenever it is
kinematically allowed to, the $\omega$ decays overwhelmingly into a
$\phi_i\phi_j$ pair (see eq.~\ref{eq:partial_widths}), thereby negating
both the aforementioned constraints.

%%%%%%%%%%%%%
%%%%%%%%%%%%%******************************************
%%%%%%%%%%%%%
\vspace*{-0.5cm}
\noindent
\subsection{Constraints on $y_e$}
\begin{figure}
\centering
\includegraphics[width =9cm,height=4.6cm]{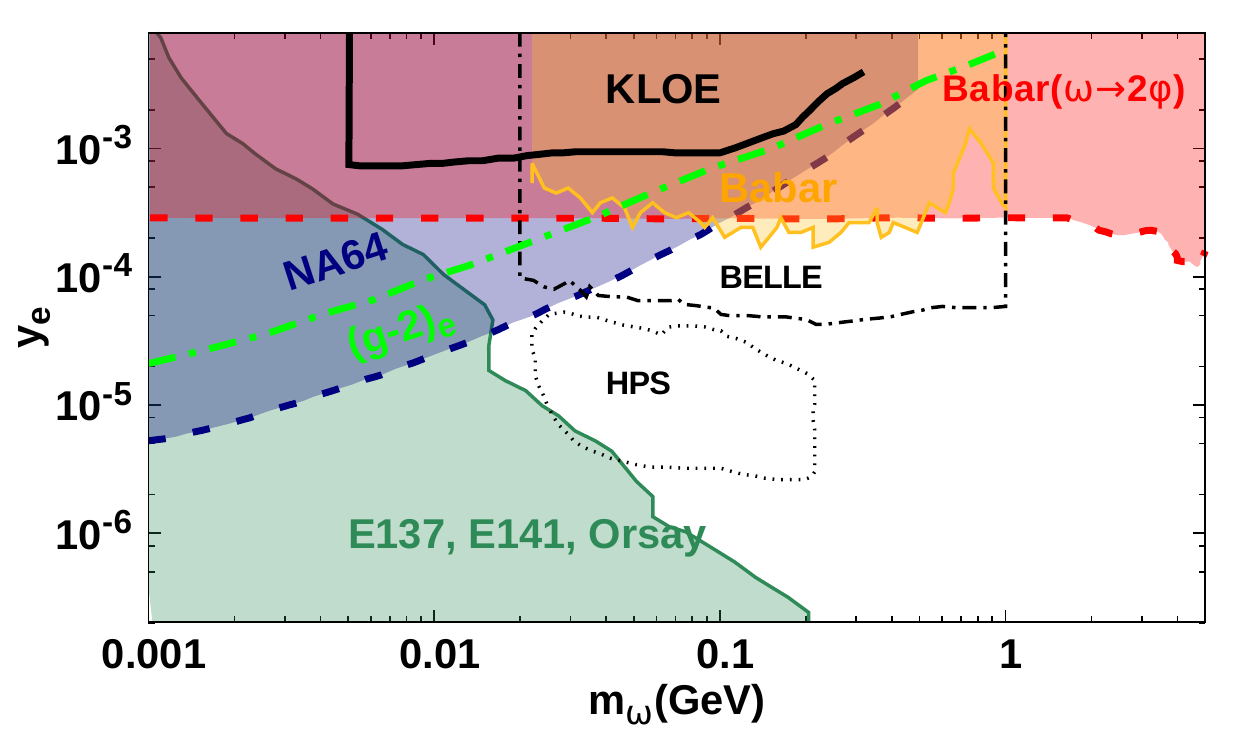}
\vskip -10pt
\caption{\em Constraints on the mediator $\omega$ coupling to electron. The
dotted curves indicate projected sensitivities from HPS~\cite{Batell:2016ove,Battaglieri:2014hga} and 
Belle-II~\cite{Batell:2016ove,Liu:2018xkx,Abe:2010gxa}.
}
\vskip -10pt
\label{elecon}
\vskip -5pt
\end{figure}

We begin by exploring the channels for $m_\omega < 2
m_\phi$ so as to remove the dependence on the invisible decay
modes. With $y_e$ being tiny, for $m_\omega\,<\,2 m_\mu$ the scalar
decay would, typically, lead to displaced vertices.  The constraints
from beam dump experiments, such as E141~\cite{Riordan:1987aw},
E137~\cite{Bjorken:1988as,Batell:2014mga} and Orsay
linac~\cite{Davier:1989wz} where $\omega$ is produced through $e^-
\,+\,N\,\to\,e^-\,+\,N\,+\,\omega$, are displayed in
Fig.\ref{elecon}. The
shape of the disallowed region is largely determined by the energy of
the decay electrons and the vertex displacement.
For $m_\omega\,>\,2m_\mu$, the small lifetime of $\omega$ drastically
reduces the sensitivity. Rather, the \Babar\ search for dark photons
via $e^+ e^- \to \gamma A' \to \gamma \ell^+
\ell^-$~\cite{Lees:2014xha} can be used to constrain
$y_e$~\cite{Batell:2016ove}.  For $y_\mu \gg y_e$, only the muonic
channel is relevant, and assuming this to be the overwhelmingly
dominant mode~\cite{Knapen:2017xzo, Liu:2018xkx} leads to strong
limits for $m_{\omega} \in[0.02,1] \gev$ (yellow region in
Fig.~\ref{elecon}). One can similarly reinterpret the \Babar\ bounds
for $m_{\omega}\, >\,1 \gev$ in terms of the scalar mediator which would
be of the same order as the sub-GeV bounds. However, we refrain from exploring that region as
it is of little interest here. The lower energy experiment
KLOE~\cite{Borodatchenkova:2005ct,Archilli:2011zc,Babusci:2012cr,Babusci:2014sta,Anastasi:2016ktq},
on the other hand, imposes a comparatively relaxed
bound~\cite{Alves:2017avw,Liu:2018xkx}.

Naturally, all the above constraints are drastically relaxed for
$m_\omega > 2m_\phi$.  Instead, $y_e$ can now be constrained from
missing energy/momentum signals.  For example, the dark photon search
of the NA64 collaboration~\cite{Banerjee:2017hhz} through
nuclei-initiated $e^- N \to e^- N A'$ with the $A'$ going invisibly,
yields constraints.  Similarly, the analogous
\Babar\ analysis~\cite{Lees:2017lec} for dark photons may be used as
well. In depicting either in Fig.~\ref{elecon}, we have, following
refs.~\cite{Batell:2016ove,Batell:2017kty,Liu:2018xkx}, interpreted
the constraints rather conservatively, eliminating a slightly larger
part of the parameter space than is strictly necessary.

The structure of eq.(\ref{eq:a_mu}) ensures that constraints from
$(g-2)_e$~\cite{Tanabashi:2018oca} are very weak.  So are those from
fifth force searches~\cite{Murata:2014nra}.  Similarly, the bounds
from the cooling of horizontal branch stars or red
giants~\cite{Hardy:2016kme} are relevant only for $m_\omega \lapp 0.1
\mev$, while those from SN1978A~\cite{Knapen:2017xzo} extend to larger
$m_\omega$ but are weaker.  Bounds from
nucleosynthesis~\cite{Knapen:2017xzo,Ghosh:2020vti} are relevant only
for $m_\omega<1\mev$ and are inapplicable in the present context.
As Fig.~\ref{elecon} (and Fig.7 of ref.\cite{Knapen:2017xzo}) shows,
for $m_\omega \gapp 0.05 \gev$, 
the $(g-2)_\mu$ favoured band of
Fig.~\ref{muoncon} is unconstrained by considerations of $y_e$ as long
as $1 \lapp n_s \lapp 10$.
%%%%%%%%%%%%%
%%%%%%%%%%%%%******************************************
%%%%%%%%%%%%%
\vspace*{-0.5cm}
\noindent
\section{Direct Detection via electron recoil: \xenon excess}
\vspace*{-0.3cm}
With the effective Yukawa couplings (\ref{eq:eff_Yuk}) in place,
the triple scalar vertices give rise to three distinct DM initiated
processes at a detector, namely $\phi_i D \to \phi_i D$ (where $D$
is a detector entity, nucleus or electron) and $\phi_2 D \to \phi_1 D$.
The former are elastic in nature with the 
typical recoil energy for an electron being
${\cal O}({\rm eV})$ and are, thus, unable
to explain the recoil-energy profile (namely, a peak at $E_{\rm
  rec.} \sim 2 \kev$) of the \xenon signal. The $g_{12}$ term in
eq.(\ref{eq:scal_lagr}), though, can lead to such events provided the
mass-splitting $\delta_m \sim {\cal O}(\kev)$.  

For an electron recoiling with energy $E$, the differential
cross-section for the atomic ionization induced by DM-electron
inelastic scattering is given by~\cite{Roberts:2019chv,Essig:2011nj}
%%%
\beq
\frac{d\langle\sigma v\rangle}{dE}
=\frac{a_{0}^{2}\bar{\sigma}_{e}}{2 m_{e}}\int_{v_{\rm min}}^{v_{\rm max}} dv
\frac{f(v)}{v}\int_{p_{-}}^{p_{+}} pdp |F_{\phi}(p)|^{2}K(E,p),
\label{eq:xenon_profile}
\eeq
%%%
where $a_{0}=1/m_{e}\alpha_{\rm em}$ and $f(v)$ is the distribution in
the DM's velocity $v$ with a Maxwellian form being a very good
approximation.  The integration limits are given by $v_{\rm min} =
\sqrt{2(E-\delta_m)/m_{2}}$ (for $E \geq \delta_m$) and $v_{\rm max} =
v_{\oplus}+v_{\rm esc}$ where $v_{\oplus}$ is the Earth's velocity and
$v_{\rm esc}$ is the local galactic escape
velocity~\cite{Lewin:1995rx}. The form factor $|F_{\phi}(p)|^{2}$, as a
function of the momentum transfer $p$, can be approximated to be unity
in the case of a heavy mediator. For free electron scattering
proceeding through $\omega$-exchange, we have
%%%
\beq
\bar{\sigma}_{e}=
y_{e}^{2} \, g_{12}^{2} \, m_{e}^{2} / ( 4 \pi m_{\omega}^{4} m_{\phi}^{2} ) \ .
\label{eq:sigmabar}
\eeq
%%%

In evaluating the integral, we use the atomic excitation factor,
$K(E,p)$ from
ref.~\cite{Roberts:2019chv}.  The integration range for $p$, as determined using
momentum conservation, is, for $E\geq\delta_m$, given by
\beq p_{\pm}=m_{2}v \pm
\sqrt{m_2^{2}v^{2}-2m_2(E-\delta_m)}\ . \label{eq:ptransfer}
\eeq
 for $\delta_{m} \sim E$, the allowed momentum transfer ranges from
 approximately zero to ${\cal O}(100\kev)$ for $m_2 \sim {\cal
   O}(100\mev)$. As ref.\cite{Harigaya:2020ckz} has pointed out,
   the second integral in eqn.(\ref{eq:xenon_profile}) would now have
   a peak at $E\approx \delta_m$. This, in turn, produces a consequent
   peak in the differential cross section. To reproduce the profile of
   the observed excess at \xenon, we need $\delta_m \sim 2 \kev$ and
   this had motivated our choice.
  
The % differential
event rate $R$ can be determined using ~\cite{Essig:2011nj} 
%%%
\beq
\frac{dR}{dE}=N_{T}\frac{\rho_{\phi_{2}}}{m_{2}} \frac{d\langle\sigma v\rangle}{dE}.
\eeq
%%%
Here, $ N_{T} \simeq 4.2 \times 10^{27}/ $tonne is the number of Xenon
atoms per unit detector mass. Since $\phi_{1,2}$ are nearly
degenerate, the energy density of incident DM particles
$\rho_{\phi_{2}} \approx \rho_{DM}/2\,\approx\, 0.15 \, {\rm
  GeV/cm^{3}}$~\cite{Hinshaw:2012aka,Ade:2015xua}.

\begin{figure}[bt]
\includegraphics[width =9cm,height=4.5cm]{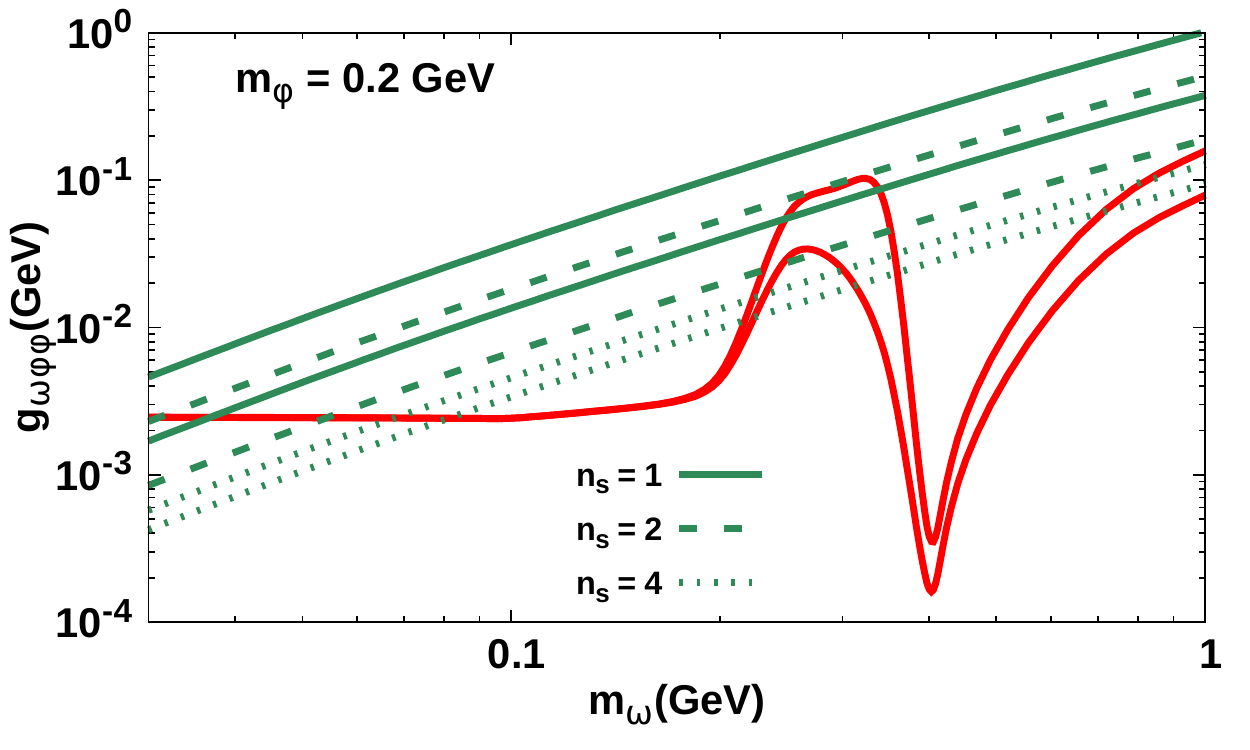}
\vskip -5pt
\includegraphics[width =9cm,height=4.5cm]{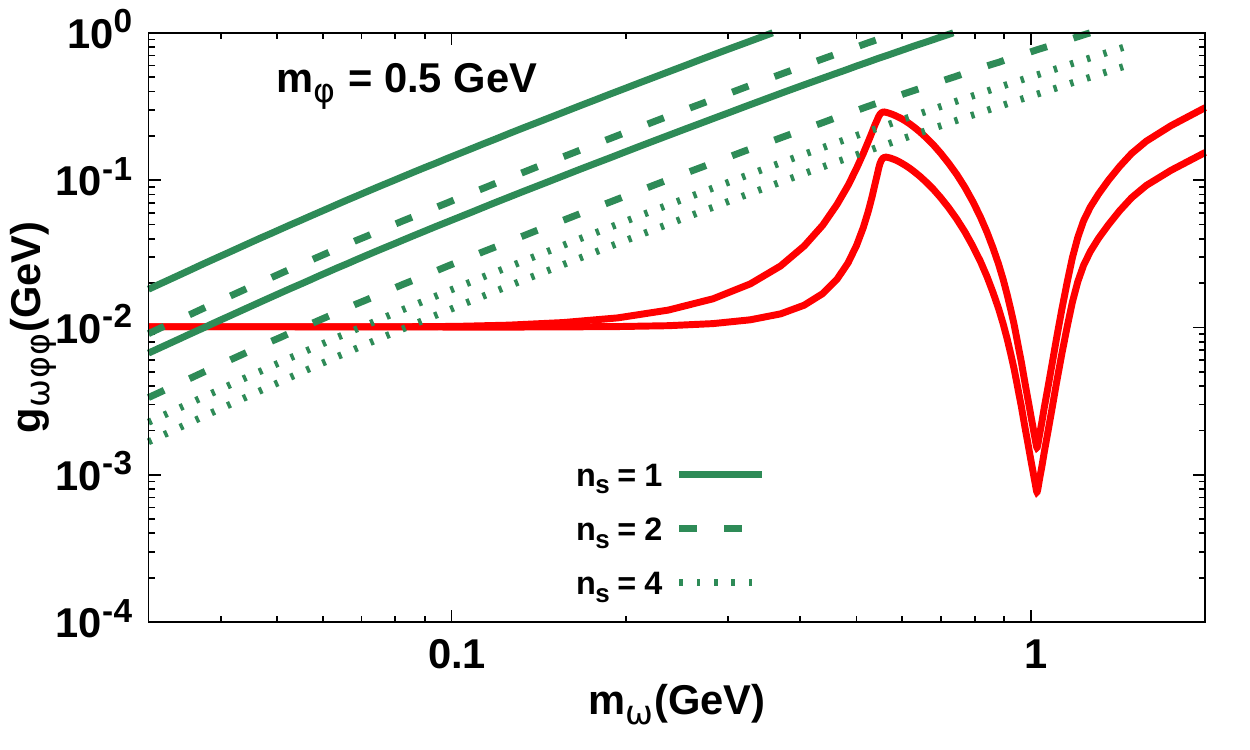}
\vskip -10pt
\caption{\em $g_{\omega \phi \phi}\,(\equiv g_{12})$ values satisfying the \xenon
excess (green) and relic density (red), as a function of $m_{\omega}$.}
\vskip -10pt
\label{gphi1}
\end{figure}

At this point we are quite well-equipped to address the \xenon
excess. With $(g-2)_{\mu}$ constraining $y_{\mu}$, a choice for $n_s$
(see eq.~\ref{eq:yuk_scaling}) determines $y_e$. This, in turn, fixes
$g_{12}$. The regions of the parameter space that can explain the
reported excess within $1\sigma$ are depicted in
Fig.~\ref{gphi1}. Note that $m_\omega \lapp 0.03$ GeV is strongly
disfavoured by low-energy data.  For a given $(m_{\omega}, m_{\phi};
y_\mu)$ combination, a larger $n_s$ would demand a smaller $g_{12}$ so
as to maintain the size of the excess, as reflected by the shifting
bands.

Apart from electrons, the DM will also scatter against the
nuclei. However, in the absence of any coupling of $\omega$ to the
quarks we only have loop-suppressed contributions to the scattering
process. This 
also invalidates the otherwise strong bound set by the {\sc cresst}
collaboration~\cite{Abdelhameed:2019hmk} for $m_{\phi} \in [0.3,1]\gev$.

%%%%%%%%%%%%%
%%%%%%%%%%%%%******************************************
%%%%%%%%%%%%%
\vspace*{-0.5cm}
\noindent
\section{Relic Abundance}
\vspace*{-0.2cm}
With $\phi_2$ having a lifetime greater than the age of the Universe,
the DM comprises equal parts of $\phi_{1,2}$. The
small $\delta_m$ ensures that the two decouple chemically well before
the heavier one could be annihilated completely or even exponentially
suppressed. By virtue of its couplings to $\phi_{1,2}$, the $\omega$
serves as a portal between the dark and the ordinary sectors.

Post decoupling, the annihilations are crucial in determining the
relic abundance.  For very light $\phi_i$, the only channel available
is $\phi_i \phi_j \to e^+ e^-$, where the two scalars could either be
the same or different. For heavier $\phi_i$, the $\mu^+ \mu^-$ and the
$\omega\omega$ modes open up. The last-mentioned, if allowed
kinematically, dominates, with propagators ($t$- and $u$-channel)
corresponding to either of $\phi_{1,2}$. Had we included a
$\omega^3$ term in the Lagrangian, a further contribution from a
$s$-channel $w$-exchange would have appeared.

Defining the yield $Y_\phi$
as the ratio of its number density and
entropy-density $s(m_\phi)$ of the universe, the relevant Boltzmann equation, 
in terms of $x \equiv m_\phi / T$ ($T$ being the temperature) and
the Hubble expansion rate $H(m_\phi)$, is
%%%
\beq
\barr{rcl}
\dis \frac{dY_{\phi_i}}{dx}& =& \dis
-\, \frac{x\,s(m_\phi)}{H(m_{\phi})}
\left[Y_{\phi_i}^2- \left(Y^{\rm equil}_{\phi_i}\right)^2\right]
\\
&& \dis \times 
\sum_{j} \left[g_{ij}^2 \sum_\ell y_\ell^2 \,\langle\sigma_{\ell\ell} v\rangle
+ \frac{1}{2} R_{ij}^2 \langle\sigma_{\omega\omega} v\rangle\right]
\earr
\eeq
%%%
where $R_{ij}
\equiv \sum_{k} g_{ik} g_{kj}$ and all the coupling constants have been
factored out of the cross-sections.  Since $\delta_m\ll T_f$, the mass
splitting has virtually no effect on the freeze-out and we have assumed
that $Y_{\phi_1} = Y_{\phi_2}$. The factor of 1/2
is occasioned by the $\omega$s being identical particles. 

To reduce the number of parameters, we make the simplifying assumption
that all three $g_{ij}$s are numerically very similar, denoting this
common value by $g_{\omega\phi\phi}$.  With $y_{\mu}$ constrained from
$(g-2)_{\mu}$, we plot, in Fig.~\ref{gphi1}, the dependence of the DM
relic abundance on $g_{\omega \phi \phi}$ as a function of the
mediator mass $m_{\omega}$ for a given DM mass. The width of the band
corresponds to the spread in $y_\mu$ (see Fig.~\ref{muoncon}).

For $m_{\omega} < m_\phi$, the processes $\phi_i \phi_j \to \omega
\omega$ are dominant. With the cross-section having only a mild
$m_\omega$-dependence, so does the requisite
$g_{\omega\phi\phi}$. Since $y_\mu$ plays only a subsidiary role, the
band collapses to virtually a single curve.  For $m_{\omega} >
m_\phi$, this channel is no more allowed and $\phi_i \phi_j \to \mu^+
\mu^-$ dominates. Consequently, $g_{\omega\phi\phi}$ must increase
with $m_\phi$ to account for the $s$-channel
suppression. Simultaneously, the allowed spread in $y_\mu$ 
becomes relevant. The strong dip around $m_\phi \sim 2 m_\omega$ is
but a consequence of resonance enhancement.

Understandably, for $n_s \lapp 10$, the relic abundance has little
dependence on it. On the other hand, the parameter space allowed by
the \xenon excess most definitely does. Consequently, it is
straightforward to identify the region of parameter space that simultaneously
explains all three viz. $(g-2)_{\mu}$, the \xenon rate and DM relic
abundance.

A larger $m_\phi$ stipulates a smaller relic number density and,
hence, a larger annihilation cross section. The requisite increase in
$g_{\omega\phi\phi}$ is not as severe as that for maintaining the
\xenon excess (see eq.~\ref{eq:sigmabar}) thereby necessitating a
larger $n_s$ for ensuring overlap (see Fig.~\ref{gphi1}).  Similarly,
$m_\omega \gg m_\phi$ would imply a progressively larger
$g_{\omega\phi\phi}$. While an overlap with the \xenon\ data can still
be achieved for $n_s < 1$, a large $g_{\omega\phi\phi}$ would result
in a large quantum correction to the scalar masses, potentially
destabilising the vacuum, unless appropriate (and allowed) terms are
added in the Lagrangian. 

\section{Other Cosmological Constraints}
\label{sec:constraints}
A DM candidate of the kind we propose can, potentially, play a
nontrivial role in the very early evolution of the universe and the
scenario is, thus, subject to constraints from observables. We, now,
consider these.

Of particular importance is the sensitivity of CMB anisotropies to
energy injection from DM annihilations into $\ell^+\ell^-$. The
relevant parameter is $p_{\rm ann} \equiv f \langle \sigma v \rangle /
m_\chi$, where $f$ is the fraction of the energy released that is
transferred to the intergalactic medium around a redshift of $\sim
600$.  The {\sc planck} collaboration~\cite{Aghanim:2018eyx} has
obtained a 95\% upper limit of 3.2 on $p^{\rm 28}_{\rm ann} \equiv
p_{\rm ann}/[10^{28} {\rm cm}^3 {\rm s}^{-1} \gev^{-1}]$. The process
responsible for the correct relic density {\em viz.} $\phi_i \phi_j
\to \omega \omega$, has $\langle\sigma v\rangle \sim 10^{-26} {\rm
  cm}^3 {\rm s}^{-1}$.  A prompt decay of $\omega$ into $e^+ e^-$
would, thus, bring this scenario into conflict with the CMB
constraints. These could, however, be evaded if the branching fraction
of $\omega$ to SM particles were less than $10^{-2} \times (m_\phi / 1
\gev)$.  This is most easily achieved by extending the dark sector. To
this end, we introduce a complex scalar $\eta$, charged under an exact
$Z_3$ symmetry (with all other fields, SM or otherwise, transforming
trivially). The most general interaction Lagrangian involving $\eta$
would, then, be
%%%%%%
\beq
\barr{rcl}
 {\cal L}_{\eta, \omega} & \ni & \dis \lambda_{\eta} |\eta|^4
  + \mu_{\omega \eta} \omega |\eta|^2 + \dis \lambda_{\omega \eta}' \omega^2 |\eta|^2
     \\[0.5ex]
 & + &  \left( \mu_{\eta} \eta^3 + \lambda_{\omega \eta} \omega \eta^3 + {\rm H.c}\right) \ .
\earr
    \label{eq:eta_lagr}         
\eeq
%%%%
While it may seem that too many free parameters have been introduced,
we shall soon see that many of them are almost irrelevant. Indeed,
$\lambda'_{\omega\eta}$ is entirely so.  Once again, we assume that
$\eta$ does not acquire a nonzero vev.

While the decay of the $\omega$ can, now, also proceed through either
of the $\mu_{\omega\eta}$ and $\lambda_{\omega\eta}$ terms, the latter
process, being a three-body decay, tends to be kinematically
suppressed. For $m_\eta$ sufficiently smaller than $m_\omega/2$, the
domination of the $\eta\eta^*$ mode over the $e^+e^-$ mode is ensured
by a moderate $\mu_{\omega\eta}$. Simultaneously, the aforementioned
CMB bounds are satisfied as long as $\mu_{\omega\eta} \gg \dis
10^{-4}m_\omega \, (m_\phi / 1\gev)^{-1/2}$.  With the $Z_3$ being
unbroken, the $\eta$ is absolutely stable and would also contribute to
the overall DM relic density. It should be noted, though, that
semi-annihilation processes such as $3 \eta \to \eta \eta^*$ (or
$\eta+2\eta^* \to 2\eta$) would occur and these play a crucial role in
determining the relic density for $\eta$ (with that for the $\eta^*$
being equal). Several diagrams contribute to each of these processes
and are listed in the Appendix. Involving both $\eta$ and $\omega$ as
mediators, the respective contributions to the amplitudes scale, in
the nonrelativistic limit, as
%%%
\vspace*{-0.1cm}
\begin{tabbing}
  $(a)\, \mu_\eta^3/m_\eta^4$, \hspace*{7em} \=
  $(b)\, \mu_{\eta} \mu_{\omega \eta}^2/m_\eta^2 (9m_{\eta}^2-m_{\omega}^2)$,\\
$(c)\, \mu_{\eta} \mu_{\omega \eta}^2/m_\eta^2 (m_{\eta}^2+m_{\omega}^2)$,\>
$(d)\, \mu_{\eta} \mu_{\omega \eta}^2/m_\eta^2 (4m_{\eta}^2-m_{\omega}^2)$,\\
$(e)\, \mu_\eta \lambda_\eta / m_\eta^2$,\>
$(f)\, \lambda_{\omega \eta} \mu_{\omega \eta}/(9m_{\eta}^2 - m_{\omega}^2)$,\\
$(g)\, \lambda_{\omega \eta} \mu_{\omega \eta}/(m_{\eta}^2 + m_{\omega}^2)$, \>
$(h)\, \lambda_{\omega \eta} \mu_{\omega \eta}/(4m_{\eta}^2-m_{\omega}^2)$\ .
\end{tabbing}
\vskip -6pt
%%%

For brevity's sake, we assume that no amplitude is
resonance-enhanced. Such an enhancement is available only for
amplitudes $(b)$ and $(f)$, and for $m_\eta \approx m_\omega / 3$
either of them is efficient enough that all other couplings can be
switched off, leaving this sector with just two, namely
$\mu_{\omega\eta}$ and one of $\mu_\eta$ and $\lambda_{\omega \eta}$.
For example, with $\mu_\eta,\mu_{\omega\eta}\sim\, 0.1\,m_\eta$
(keeping others parameters zero), we get $\Omega_\eta
h^2\,\sim\,10^{-4}$~\cite{Bhattacharya:2020}. Similarly, for $\mu_\eta
= 0$ and $\lambda_{\omega\eta}\sim 0.1 \, (m_\eta/1\gev)$, one obtains
$\Omega_\eta h^2\,\sim\,10^{-3}$.  Even far away from such a
resonance, a combination of ${\rm max}(\mu_\eta,\mu_{\omega\eta})
\gtrsim {\cal O}(5\,m_\eta)$ and $\lambda_\eta, \lambda_{\omega\eta}
\gapp 0.01$ suppresses the relic $\eta$-density to ${\cal
  O}(10^{-3})$.  This suppression suffices to ensure that the sizable
$\eta\eta^{(*)}$ self-interaction that such terms engender are
consistent with the constraints from the Bullet cluster~\cite{Randall:2008}.

While it might seem that the $\eta\eta^* \to e^+ e^-$ process
  would resurrect the problem with the CMB, it is not so. Even though
  each $\omega$ decay would create multiple $\eta$-particles, note
  that the smallness of $y_e$ ensures that the $\eta\eta^* \to e^+
  e^-$ cross section is much smaller than that for $3\eta \to 2\eta$,
  despite the latter being a $3 \to 2$ process. Thus, the $\eta$s
  settle to the tiny relic density much faster than they pump energy
  into the CMB. In addition, the longer injection time further
  ameliorates the problem. Note that this argument holds as long as
  $\mu_{\omega\eta}$ is not too large ($\lapp 10\mev$), almost
  independent of whether we are close to the resonance region.

Processes such as $\phi_ie^-\,\to\,\phi_je^-$ maintain kinetic
equilibrium and keep the dark sector in thermal contact with the
plasma until $T_{\rm kin}$, when it decouples.  For $n_s \sim 4$,
comparing the interaction rate to the expansion (Hubble) rate gives
$T_{\rm kin} \sim {\rm MeV}$.  After the decoupling, the DM is no
longer in kinetic equilibrium with the SM thermal bath and begins to
cool more rapidly.

The inter-conversion process $\phi_2\phi_2\,\to\,\phi_1\phi_1$,
nonetheless, continues to be efficient until the temperature of the
dark sector falls below $T'< T_{\rm kin}$. If $T'< \delta_m$, the
fractional abundance of $\phi_2$ would be exponentially suppressed,
with $N_2/N_1\sim e^{-\delta_m/T'}$. Using the formalism of
refs.\cite{Finkbeiner:2009mi,Batell:2009vb}, we find, though, that $T'
\gapp \mathcal{O}(100 \kev)$ and $N_2/N_1\sim\,1$. Similarly,
  the interconversion as well as the scattering rates are much smaller
  than the constraints from structure formation.
%%%%%%%%%%%%%
%%%%%%%%%%%%%******************************************
%%%%%%%%%%%%%
\vspace*{-0.18cm}
\section{Summary and Outlook}
\vspace*{-0.2cm} We present a very economical model that
simultaneously explains the \xenon excess (through inelastic DM
scattering), as well as the anomalous magnetic moment of the muon
while producing the requisite dark matter relic density. A single
leptophilic scalar $\omega$, lighter than the DM, generates the
requisite $a_\mu$ while serving as a portal between the dark and the
visible sectors. A tiny mass-splitting of ${\cal O}(\kev)$ between the
two components of the scalar field representing the DM is engendered
by a soft term in the scalar potential (or, potentially, by a nonzero
$\langle \omega \rangle$). The very smallness of the splitting allows
the heavier DM component to be stable on cosmological time
scales. While the DM mass is required to be relatively small,
viz. ${\cal O}(100\mev)$, a sufficient parameter space exists
satisfying all constraints, experimental (beam dumps, colliders etc)
astrophysical (stellar cooling) and cosmological (BBN, $N_{\rm
  eff}$). Constraints arising from energy injections into the CMBR are
evaded by introducing a third scalar field $\eta$ to which the
mediator $\omega$ dominantly decays into. While $\eta$ itself is
cosmologically stable, its interactions drive its relic density to
less than ${\cal O}(10^{-3})$.

  Competing constraints render
  the model eminently testable and, thus, interesting. For example, in
  Fig.~\ref{elecon} we have indicated the projected sensitivities for
  $y_{e}$ from Belle-II~\cite{Batell:2016ove,Liu:2018xkx,Abe:2010gxa}
  and the Heavy Photon Search (HPS)
  experiments~\cite{Batell:2016ove,Battaglieri:2014hga}. A similar
  Belle-II projection for $y_{\mu}$~\cite{Batell:2017kty} has been
  indicated in Fig.\ref{muoncon}. The FASER
  experiment~\cite{PhysRevD.99.095011} too can probe such parameters.
  Clearly, a very large part of the favoured parameter space would be
  testable in the near future both in terrestrial experiments as well
  as CMBR observations.  Also worth studying are the consequences of a
  nonzero $\langle \omega \rangle$, especially in the context of
  finite-temperature corrections, for this presents intriguing
  possibilities as far as cosmological history is concerned, whether
  it be in terms of phase transitions, small late stage inflation
  etc. We hope to return to such issues at a later date.
\vspace*{-0.6cm}
\section*{Acknowledgement}
\vspace*{-0.3cm}
We thank Filippo Sala for constructive criticism and Abhijit Kumar
Saha for bringing to our notice Chandra and XMM-Newton observations. For partial
support, DS acknowledges a Ramanujan Fellowship grant of DST, India, 
VS thanks the UGC, India while DC and SM acknowledge research grant 
CRG/2018/004889 of the SERB, India.

%\vspace{0.5cm}

\section*{Appendix}
\label{sec:appendix}

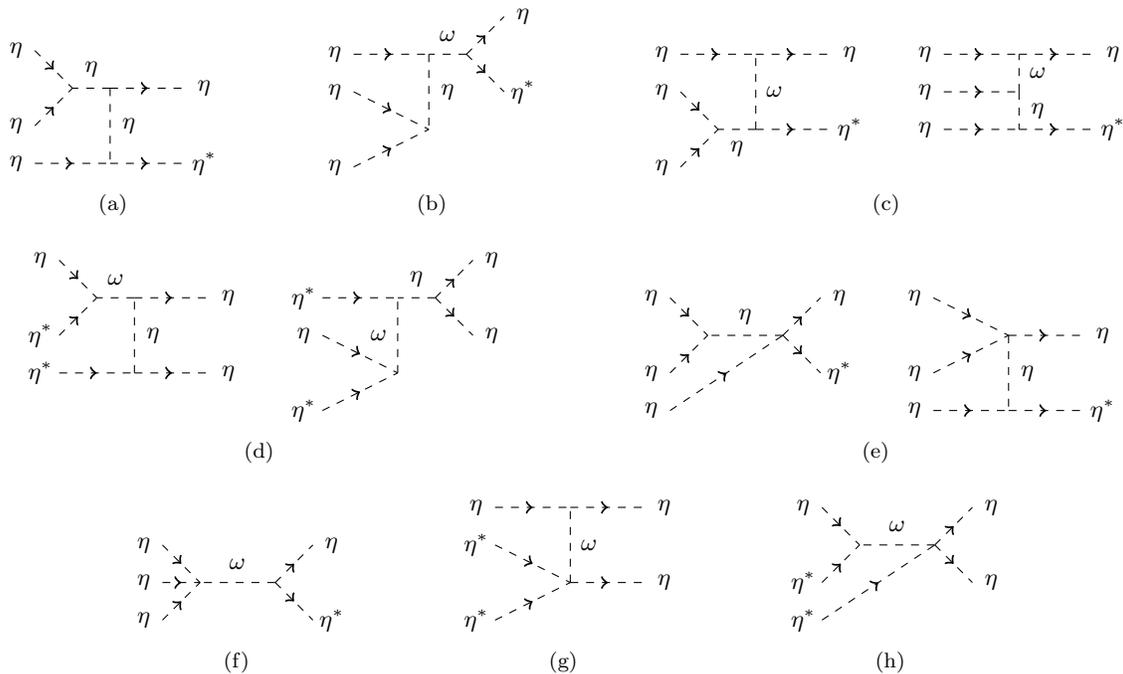
\begin{figure}[!h]
\subfigure[]{%
\begin{tikzpicture}
\draw[dashed,black] (0,0) --(-0.5,0);
\draw[scalar,black] (0,0) --(1,0);
\draw[scalar,black] (-1,0.5) --(-0.5,0);
\draw[scalar,black] (-1,-0.5) --(-0.5,0);
\draw[dashed,black] (0,0) --(0,-1);
\draw[scalar,black] (-1,-1) --(0,-1);
\draw[scalar,black] (0,-1) --(1,-1);

\node at (-1.25,0.5) {$\eta$};
\node at (-1.25,-0.5) {$\eta$};
\node at (-1.25,-1) {$\eta$};
\node at (1.25,0) {$\eta$};
\node at (1.25,-1) {$\eta^*$};
\node at (-0.25,0.25) {$\eta$};
\node at (0.25,-0.5) {$\eta$};
\end{tikzpicture}}  
\hspace*{30pt}
\subfigure[]{%
\begin{tikzpicture}
\draw[scalar,black] (-1,0) --(0,0);
\draw[dashed,black] (0.5,0) --(0,0);
\draw[scalar,black] (0.5,0) --(1,0.5);
\draw[scalar,black] (0.5,0) --(1,-0.5);
\draw[dashed,black] (0,0) --(0,-1);
\draw[scalar,black] (-1,-0.5) --(0,-1);
\draw[scalar,black] (-1,-1.5) --(0,-1);

\node at (-1.25,0) {$\eta$};
\node at (-1.25,-0.5) {$\eta$};
\node at (-1.25,-1.5) {$\eta$};
\node at (1.25,0.5) {$\eta$};
\node at (1.25,-0.5) {$\eta^*$};
\node at (0.25,0.25) {$\omega$};
\node at (0.25,-0.5) {$\eta$};
\end{tikzpicture}} 
\hspace*{30pt}
\subfigure[]{
\begin{tikzpicture}
\draw[scalar,black] (-1,0) --(0,0);
\draw[scalar,black] (0,0) --(1,0);
\draw[dashed,black] (0,-1) --(0,0);
\draw[dashed,black] (0,-1) --(-0.5,-1);
\draw[scalar,black] (0,-1) --(1,-1);
\draw[scalar,black] (-1,-0.5) --(-0.5,-1);
\draw[scalar,black] (-1,-1.5) --(-0.5,-1);

\node at (-1.25,0) {$\eta$};
\node at (-1.25,-0.5) {$\eta$};
\node at (-1.25,-1.5) {$\eta$};
\node at (1.25,0) {$\eta$};
\node at (1.25,-1) {$\eta^*$};
\node at (0.25,-0.5) {$\omega$};
\node at (-0.25,-1.25) {$\eta$};

\draw[scalar,black] (2.5,0) --(3.5,0);
\draw[scalar,black] (3.5,0) --(4.5,0);
\draw[dashed,black] (3.5,-0.5) --(3.5,0);
\draw[dashed,black] (3.5,-0.5) --(3.5,-1);
\draw[scalar,black] (2.5,-0.5) --(3.5,-0.5);
\draw[scalar,black] (2.5,-1) --(3.5,-1);
\draw[scalar,black] (3.5,-1) --(4.5,-1);

\node at (2.25,0) {$\eta$};
\node at (2.25,-0.5) {$\eta$};
\node at (2.25,-1) {$\eta$};
\node at (4.75,0) {$\eta$};
\node at (4.75,-1) {$\eta^*$}; 
\node at (3.75,-0.25) {$\omega$};
\node at (3.75,-0.75) {$\eta$}; 
\end{tikzpicture}}
\subfigure[]{
\begin{tikzpicture}
\draw[dashed,black] (0,0) --(-0.5,0);
\draw[scalar,black] (0,0) --(1,0);
\draw[scalar,black] (-1,0.5) --(-0.5,0);
\draw[scalar,black] (-1,-0.5) --(-0.5,0);
\draw[dashed,black] (0,-1) --(0,0);
\draw[scalar,black] (-1,-1) --(0,-1);
\draw[scalar,black] (0,-1) --(1,-1);

\node at (-1.25,0.5) {$\eta$};
\node at (-1.25,-0.5) {$\eta^*$};
\node at (-1.25,-1) {$\eta^*$};
\node at (1.25,0) {$\eta$};
\node at (1.25,-1) {$\eta$};
\node at (-0.25,0.25) {$\omega$};
\node at (0.25,-0.5) {$\eta$};

\draw[scalar,black] (2.5,0) --(3.5,0);
\draw[dashed,black] (4,0) --(3.5,0);
\draw[scalar,black] (4,0) --(4.5,0.5);
\draw[scalar,black] (4,0) --(4.5,-0.5);
\draw[dashed,black] (3.5,-1) --(3.5,0);
\draw[scalar,black] (2.5,-0.5) --(3.5,-1);
\draw[scalar,black] (2.5,-1.5) --(3.5,-1);

\node at (2.25,0) {$\eta^*$};
\node at (2.25,-0.5) {$\eta$};
\node at (2.25,-1.5) {$\eta^*$};
\node at (4.75,0.5) {$\eta$};
\node at (4.75,-0.5) {$\eta$};
\node at (3.75,0.25) {$\eta$};
\node at (3.25,-0.5) {$\omega$};
\end{tikzpicture}} 
\hspace*{40pt}
\subfigure[]{
\begin{tikzpicture}
\draw[dashed,black] (0.5,0) --(-0.5,0);
\draw[scalar,black] (-1,0.5) --(-0.5,0);
\draw[scalar,black] (-1,-0.5) --(-0.5,0);
\draw[scalar,black] (-1,-1) --(0.5,0);
\draw[scalar,black] (0.5,0) --(1,0.5);
\draw[scalar,black] (0.5,0) --(1,-0.5);

\node at (-1.25,0.5) {$\eta$};
\node at (-1.25,-0.5) {$\eta$};
\node at (-1.25,-1) {$\eta$};
\node at (1.25,0.5) {$\eta$};
\node at (1.25,-0.5) {$\eta^*$};
\node at (0,0.25) {$\eta$};

\draw[scalar,black] (3.5,0) --(4.5,0);
\draw[scalar,black] (2.5,0.5) --(3.5,0);
\draw[scalar,black] (2.5,-0.5) --(3.5,0);
\draw[dashed,black] (3.5,0) --(3.5,-1);
\draw[scalar,black] (2.5,-1) --(3.5,-1);
\draw[scalar,black] (3.5,-1) --(4.5,-1);

\node at (2.25,0.5) {$\eta$};
\node at (2.25,-0.5) {$\eta$};
\node at (2.25,-1) {$\eta$};
\node at (4.75,0) {$\eta$};
\node at (4.75,-1) {$\eta^*$};
\node at (3.75,-0.5) {$\eta$};
\end{tikzpicture}} \\
\subfigure[]{
\begin{tikzpicture}
\draw[scalar,black] (-1,-0.5) --(-0.5,-0.5);
\draw[dashed,black] (0.5,-0.5) --(-0.5,-0.5);
\draw[scalar,black] (-1,0) --(-0.5,-0.5);
\draw[scalar,black] (-1,-1) --(-0.5,-0.5);
\draw[scalar,black] (0.5,-0.5) --(1,0);
\draw[scalar,black] (0.5,-0.5) --(1,-1);

\node at (-1.25,-0) {$\eta$};
\node at (-1.25,-0.5) {$\eta$};
\node at (-1.25,-1) {$\eta$};
\node at (1.25,0) {$\eta$};
\node at (1.25,-1) {$\eta^*$};
\node at (0,-0.25) {$\omega$};
\end{tikzpicture}} 
\hspace*{30pt}
\subfigure[]{
\begin{tikzpicture}
\draw[scalar,black] (-1,0) --(0,0);
\draw[scalar,black] (0,0) --(1,0);
\draw[dashed,black] (0,-1) --(0,0);
\draw[scalar,black] (0,-1) --(1,-1);
\draw[scalar,black] (-1,-0.5) --(0,-1);
\draw[scalar,black] (-1,-1.5) --(0,-1);

\node at (-1.25,0) {$\eta$};
\node at (-1.25,-0.5) {$\eta^*$};
\node at (-1.25,-1.5) {$\eta^*$};
\node at (1.25,0) {$\eta$};
\node at (1.25,-1) {$\eta$};
\node at (0.25,-0.5) {$\omega$};
\end{tikzpicture}} 
\hspace*{30pt}
\subfigure[]{
\begin{tikzpicture}
\draw[dashed,black] (0.5,0) --(-0.5,0);
\draw[scalar,black] (-1,0.5) --(-0.5,0);
\draw[scalar,black] (-1,-0.5) --(-0.5,0);
\draw[scalar,black] (0.5,0) --(1,0.5);
\draw[scalar,black] (0.5,0) --(1,-0.5);
\draw[scalar,black] (-1,-1) --(0.5,0);

\node at (-1.25,0.5) {$\eta$};
\node at (-1.25,-0.5) {$\eta^*$};
\node at (-1.25,-1) {$\eta^*$};
\node at (1.25,0.5) {$\eta$};
\node at (1.25,-0.5) {$\eta$};
\node at (0,0.25) {$\omega$};
\end{tikzpicture}}
  \caption{Feynman diagrams corresponding to various sub-processes leading to 3-to-2 annihilation of the complex scalar, $\eta$ , with arrows representing the momentum flow.}
\label{fig:diagrams}
\end{figure}

The amplitudes in sec.~\ref{sec:constraints} correspond to various diagrams as shown in fig.~\ref{fig:diagrams}, respectively. The diagrams categorically belong to either of the two processes $viz.$ $3 \eta \to \eta \eta^*$ or $\eta+2\eta^* \to 2\eta$.
%$$\eta (p_1)~\eta (p_2)~\eta (p_3)~\to~\eta (p_4)~\eta^* (p_5)$$
%$$\eta^* (p_1)~\eta (p_2)~\eta^* (p_3)~\to~\eta (p_4)~\eta (p_5)$$.
\onecolumngrid

\twocolumngrid

\bibliographystyle{apsrev4-1}
\bibliography{reference}
\end{document}